\documentstyle[psfig,11pt]{article}

\def\etal{{\it et al.}}
\def\xb{4U1915-05}

\def\ergs{ergs s$^{-1}$}

\begin{document}
\vspace{1.0cm}

{\Large \bf LOW AND HIGH FREQUENCY QUASI-PERIODIC OSCILLATIONS IN \xb}

\vspace{1.0cm}

L. Boirin$^1$, D. Barret$^1$, J. F. Olive$^1$, J. E. Grindlay$^2$ and P.F. Bloser$^{2}$

\vspace{1.0cm}
$^1${\it CESR, 9 Avenue du Colonel Roche, 31028 Toulouse
Cedex 04, France (Laurence.Boirin@cesr.fr)}\\ $^2${\it Harvard
Smithsonian Center for Astrophysics, 60 Garden Street, Cambridge, MA
02138, USA} \\
\vspace{0.5cm}

%%%%%%%%%%%%%%%%%%%%
\section*{ABSTRACT}
%%%%%%%%%%%%%%%%%%%%
The burster and dipper Low Mass X-ray Binary (LMXB) \xb~ (also known
as XB1916-053) was observed by the {\it Rossi X-ray Timing Explorer}
(RXTE) 19 times for a total exposure of 140 ks between 1996 February
and October. Here we report on the discovery of Low Frequency (10-40
Hz) Quasi-Periodic Oscillations (LFQPOs) from \xb.  The properties of
the LFQPOs are related to the presence of the High Frequency QPOs
(HFQPOs) detected simultaneously.  We have observed a correlation
between the LFQPO frequency and source count rate, as well as a
correlation (linear) between the LFQPOs and HFQPOs.  Both results
cannot be explained by a Beat Frequency Model (BFM). They are also
hardly compatible with predictions from the inner disk precession
model.

%%%%%%%%%%%%%%%%%%%%%%%
\section{INTRODUCTION}
%%%%%%%%%%%%%%%%%%%%%%%

Prior to RXTE, GINGA observations had shown that \xb~displayed very
low variability at frequencies below 1 Hz (Yoshida, 1992).  The same
observations suggested also that \xb~could be an Atoll source. \xb~was
observed by RXTE both in a high intensity/soft state (${\rm
L}_{2-20\;\rm{keV}}\sim1.4 \times 10^{37}$ \ergs, 9.3 kpc) and a low
intensity/hard state (${\rm L}_{2-20\;\rm{keV}}\sim3.2 \times 10^{36}$
\ergs). A correlated timing and spectral study confirmed the Atoll
nature of the source (Boirin \etal, 1999). In the low intensity
regime, HFQPOs were detected between 600 and 1000 Hz (Barret \etal,
1999, Boirin \etal, 1999).  Their frequency positively correlates with
the count rate, except for an observation performed on October 29th
when a 600 Hz QPO was detected.  For two observations, a second
marginally significant signal was found at a frequency 350 Hz lower
than the main HFQPO.  Applying the ``shift and add'' technique (Mendez
\etal, 1998) revealed twin HFQPOs separated by 355 Hz at 650 and 1005
Hz. In this paper we report on the discovery of LFQPOs in \xb, and
relate their properties to those of HFQPOs.

%%%%%%%%%%%%%%%%%%%%%%%
\section{DISCOVERY OF LFQPOs}
%%%%%%%%%%%%%%%%%%%%%%%

\begin{figure}[!b]
\centerline{\psfig{figure=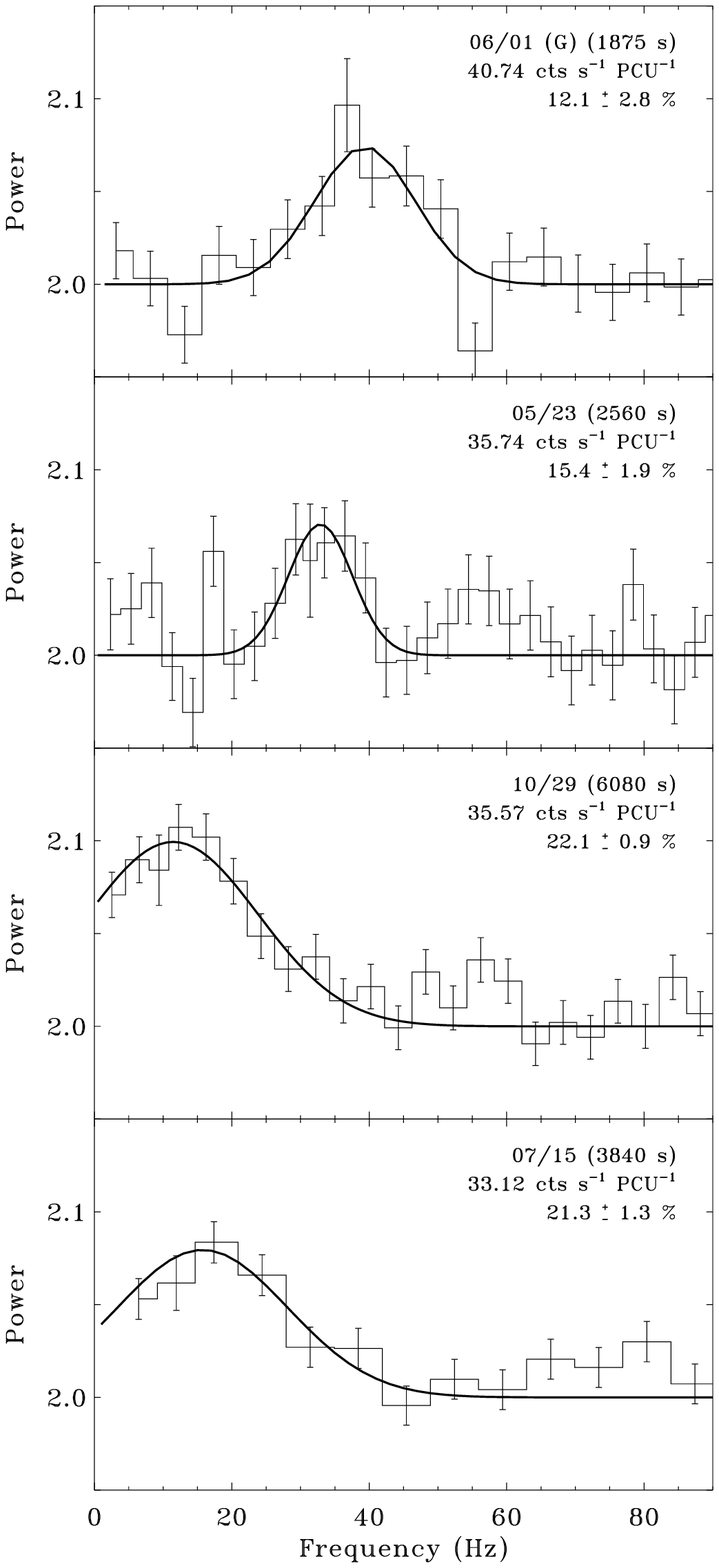,width=9cm}\psfig{figure=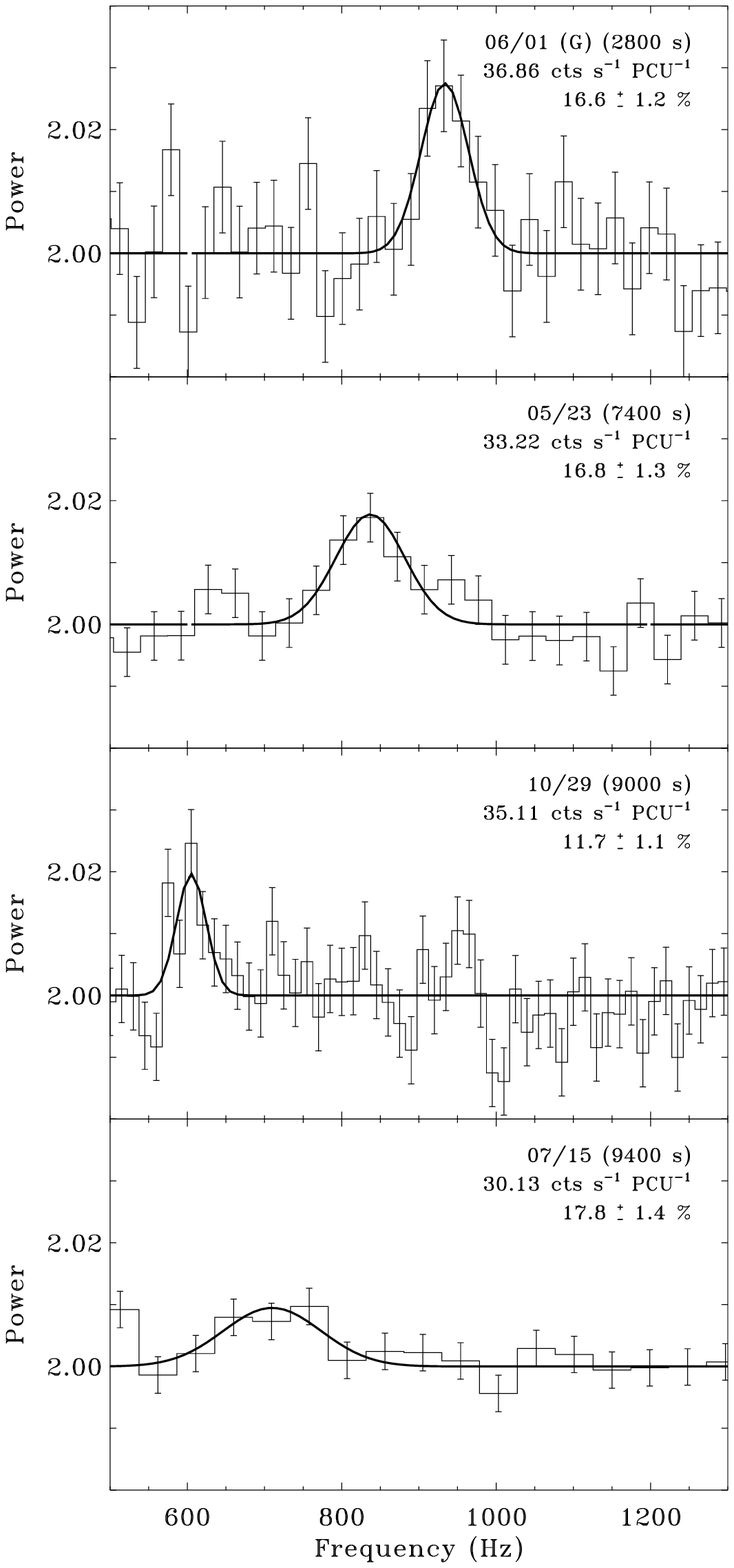,width=9cm}}
\caption{PDS for the observations displaying both LFQPOs 
(left) and HFQPOs (right). 
The observation date, exposure time, 5-30 keV count rate before background 
subtraction and RMS are indicated. For the LFQPOs, the RMS is computed in the 
$3\times10^{-3}-100$ Hz range; the dipping parts of the
light curves were filtered out.}
\label{qpo}
\end{figure}

\begin{figure}[!t]
\vspace{-1.5cm}
\centerline{\psfig{figure=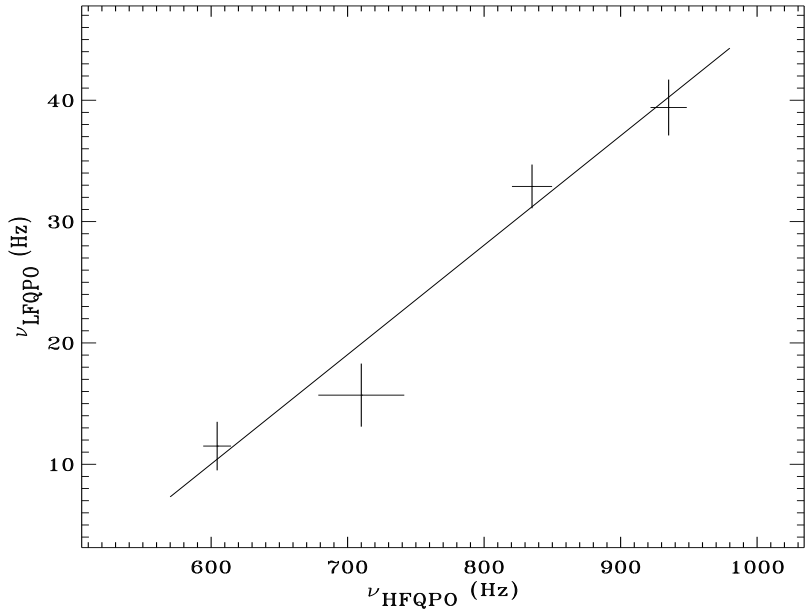,width=16cm,height=12cm}}
\vspace{-1.5cm}
\caption{LFQPO versus HFQPO frequency and linear fit.}
\label{correlation}
\end{figure}

We used the high time resolution Proportional Counter Array (PCA) data
(122 $\mu$s resolution and 0.95 $\mu$s for the GoodXenon (G) mode).
Power Density Spectra (PDS) ($3\times10^{-3}-100 $ Hz) were computed
in the 5-30 keV range (for all details about the observation and the
data analysis, see Boirin \etal, 1999).  The source displays very low
variability below 1 Hz.  The Root Mean Square (RMS) integrated up to 1
Hz ranges between $\sim$ 3 and 6 \% independently of the
spectral/intensity state.  Above 1 Hz, in the high intensity state,
\xb~ does not display any variability either. On the other hand, at
lower intensities, the source displays larger variability, with RMS
ranging from roughly 8 to 28 \% as the count rate decreases.  A
careful inspection of the LF PDS revealed the presence of broad
QPO-like features.  Figure \ref{qpo} (left) shows the LFQPOs in order
of decreasing count rate from top to bottom. Parameters of the LFQPOs
are given on the plot (see also Table \ref{thirring}). With the
exception of the October 29th observation, there is a clear
correlation between the LFQPOs and count rate. LFQPOs were detected in
five observations. In four of them, HFQPOs were detected
simultaneously (above the 5 $\sigma$ level) (Barret \etal, 1999, see
Figure \ref{qpo}). Figure \ref{correlation} is a plot of the LFQPO
frequency against the HFQPO frequency.

\section{DISCUSSION}

The main results of this paper can be summarized as follows: we have
observed a positive correlation of the LFQPOs with the count rate as
well as a correlation between the LFQPOs and HFQPOs. \xb~ is the third
Atoll source, to date, to display such a behaviour; the other two
being 4U0614+091 (Ford, 1997) and 4U1728-34 (Ford \& van der Klis,
1998).  Such correlations are already well known in Z sources (eg Sco
X-1, van der Klis \etal, 1997), thus suggesting that similar processes
are at work in both kinds of LMXBs. The origin of LFQPOs in Atoll
sources is unclear. The magnetospheric BFM (Alpar \etal, 1985) long
used to explain the Horizontal Branch (HB) QPOs in Z sources cannot be
safely extrapolated to Atoll sources, whose magnetic field and
accretion rate are much lower than the inferred value for Z
sources. Furthermore, the BFM, if used to explain the LFQPOs cannot
account at the same time for the presence of HFQPOs (van der Klis
\etal, 1997). Note also that in any of such models one would expect a
strict correlation between the LFQPOs and count rate. As said earlier,
the October 29th observation departs from such a correlation.  It is
interesting to note that the 600 Hz HFQPO is also outside the
correlation between HFQPOs and count rate, but surprisingly (as shown
in Figure \ref{correlation}) falls right on the correlation between
LFQPOs and HFQPOs.

Stella and Vietri (1998) developed the Lense-Thirring (LT) inner disk
precession model that might account for LFQPOs when HFQPOs are
present. This model has been succesfully applied to three sources
(4U1728-34, 4U0614+091 and KS1731-260, Stella \& Vietri, 1998) but
might have some problems. First, in 4U1735-44, Wijnands \etal~(1998)
computed a precession frequency greater by a factor of 2 than the
LFQPO observed.  Second, in 4U1728-34, although the LFQPO frequency
varies as the squared of the HFQPO frequency as expected for the
precession frequency, the parameters derived for the Neutron Star (NS)
are not allowed by classical models (Ford \etal, 1998). Finally, a
similar conclusion was also reached when the LT model was tried on Z
sources (Stella \& Vietri, 1998, Jonker \etal, 1998).  For \xb, we
assumed a NS spin frequency of 355 Hz and the keplerian frequency
equal to the observed HFQPO frequency.  We used the same parameters as
Stella and Vietri for the NS (NS mass ${\rm M}_{\rm o } = 1.97 {\rm
M}_{\odot}$, moment of inertia ${\rm I} = 1.98{\rm M}_{\rm o }10^{45}$
g cm$^2$).  The results are listed in Table \ref{thirring}.
\begin{table}[!t]
\begin{center}
\begin{tabular}{|llll|}
\hline
Date &  $\nu_{\;HFQPO}$ & $\nu_{\;LFQPO}$ & $ \nu_{\;precession}$  \\
     & (Hz)& (Hz)& (Hz)\\
\hline
\hline
05/23  & 835.1$^{+14.7}_{-14.2}$ & $32.9\pm1.8$ &  18.1\\
\hline
06/01 (G) & $935.2\pm13.3$ & $39.4\pm2.3$ & 22.5\\ 
\hline
07/15  & 710.0$^{+31.5}_{-30.7}$ & $15.7\pm2.6$ & 13.2 \\
\hline
10/29  & 604.4$^{+9.8}_{-10.2}$ & $10.7\pm1.5$ & 9.6\\
       & 604.4 + 355  & & 23.7 \\
\hline
\end{tabular}
\end{center}
\vspace{-5mm}
\caption{Parameters of High and Low Frequency QPOs detected
  simultaneously from \xb. $\nu_{precession}$ is the expected frequency 
 of the inner-disk precession model (Stella \& Vietri, 1998) assuming
 that \xb~ contains a 355 Hz rotating NS.}
\label{thirring}
\end{table}
The precession frequency matches the LFQPO frequency for the July 15th
observation. For the October 29th, it would be also the case if the
$\sim$ 600 Hz HFQPO were associated with the inner-disk keplerian
frequency.  For the remaining two observations (May 23rd and June
1st), the precession frequency is lower by about a factor of $\sim$ 2
than the QPO frequency observed. Since the NS parameters used are
supposed to yield precession frequencies close to the maximum values
allowed in classical NS models, it is unlikely that the LT model can
be pushed up to match our LFQPO.  Note however, that for the latter
two observations, the LFQPO frequency would roughly match the
precession frequency if the HFQPO observed were the lowest of a twin
pair separated by 355 Hz, or the highest but with a NS spin at $\sim 2
\times 355$ Hz.  In any case, the LFQPO does not seem to follow the
quadratic dependency on the HFQPO, predicted in the LT interpretation
(Figure \ref{correlation}).  Excluding the peculiar October 29th
observation, a fit with a powerlaw gives an index of 3.4 (reduced
$\chi^2=3.5$). For indication, a simple linear fit yields a lower
reduced $\chi^2$ (1.3).

\section{CONCLUSIONS}

Our study of the timing properties led to the discovery of LFQPOs. It
further revealed a positive correlation between the LFQPOs and count
rate, and a correlation between the LFQPOs and HFQPOs. We have shown
that our results are hardly compatible with models that have been put
forward to explain LFQPOs in Atoll sources. In any case, more
observations are needed to better constrain the dependency rules of
LFQPOs versus HFQPOs, which in turn might put more stringent
constraints on the models.

We thank J. Swank and A. Smale for their support to this project, as
well the RXTE/GOF team for his assistance in the data reduction and
analysis.

\section{REFERENCES}

\begin{itemize}
\setlength{\itemindent}{-8mm}
\setlength{\itemsep}{-1mm}
\item [] {Alpar}, M.A., \& {Shaham}, J., 1985, Nature, {\bf316}, 239
\item [] {Barret}, D., \etal, 1999, Astr. Lett. and Comm., in press
\item [] {Boirin}, L., \etal, 1999, {\em A\&A}, in preparation
\item [] {Ford}, E.~C., 1997, {PhD Thesis}, Columbia University
\item [] {Ford}, E.~C., \& {Van Der Klis}, M., 1998, {\em ApJL}, {\bf 506}, L39
\item [] {Jonker}, P.G., \etal, 1998,{\em ApJL}, {\bf 499}, L191
\item [] {Mendez}, M., \etal, 1998, {\em ApJL}, {\bf 494}, L65
\item [] {Stella}, L., \& Vietri, M., 1998, {\em ApJL}, {\bf 492}, L59
\item [] {Van der Klis}, M., Wijnands, R., Horne, K. and Chen, W.,
  1997, {\em ApJL}, {\bf 481}, L97
\item [] {Wijnands}, R., \etal, 1997, {\em ApJL}, {\bf 490}, L157
\item [] {Wijnands}, R., \etal, 1998, {\em ApJL}, {\bf495}, L39
\item [] {Yoshida}, K., 1992, {PhD Thesis}, Tokyo University

\end{itemize}

\end{document}